\documentclass[conference]{IEEEtran}
\usepackage{graphicx}
\usepackage{epsf}
\usepackage{float}
\usepackage{stfloats}
\usepackage{subcaption}
\usepackage{amsmath}
\usepackage{amsfonts}
\usepackage{pdfpages}
  \usepackage{caption}
\usepackage{algpseudocode}
\usepackage{multicol}
\usepackage{cite}
\usepackage[ruled,linesnumbered]{algorithm2e}
\ifCLASSINFOpdf
\else
\fi
\begin{document}
\title{A Feature-Based Bayesian Method for Content Popularity Prediction in Edge-Caching Networks}
\author{\IEEEauthorblockN{Sajad Mehrizi,
		Anestis Tsakmalis,
		Symeon Chatzinotas, 
		Bj{\"o}rn Ottersten }
	\IEEEauthorblockA{Interdisciplinary Centre for Security, Reliability and Trust (SnT), University of Luxembourg
	}
	{$\left\{ sajad.mehrizi, anestis.tsakmalis, symeon.chatzinotas, bjorn.ottersten \right\}$$@uni.lu$ }
}
\maketitle
\begin{abstract}
Edge-caching is recognized as an efficient technique for future wireless cellular networks to improve  network capacity and  user-perceived quality of experience.  Due to the random content requests and the limited  cache memory, designing an efficient caching policy is a challenge. To enhance the performance of caching systems, an accurate content request  prediction algorithm is essential. Here, we introduce a flexible model, a Poisson regressor based on a Gaussian process, for the content request distribution in stationary environments.  Our proposed model can incorporate the content features as  side information for prediction enhancement. In order to learn the model parameters, which yield the Poisson rates or alternatively content \textit{popularities}, we invoke the Bayesian approach which is very robust against over-fitting.
 However,  the posterior distribution in the Bayes formula is  analytically intractable to compute. To tackle this issue, we apply a Monte Carlo Markov Chain (MCMC) method to approximate the posterior distribution.  Two types of predictive distributions are formulated for the requests of  existing contents and for the requests of a newly-added content. Finally, simulation
 results are provided to confirm the accuracy of the developed
 content popularity learning approach.
\end{abstract}
\section{Introduction}
 Mobile data traffic is forecast to increase at a $47 \%$ compound annual growth rate (CAGR) from 2016 to 2021, two times faster than the growth of global IP fixed traffic during the same period.~\cite{indexglobal}. This is largely due to the growth in both the number of mobile devices and the user interest towards high-rate multimedia applications. Nevertheless, supporting such a huge data traffic turns to be a big challenge  which indicates the need for developing new architectures. To mitigate this issue, edge-caching is recognized as one of the leading technologies~\cite{shanmugam2013femtocaching,ge20145g}. It can bring the requested content from the core network close to the end mobile user, instead of downloading the same content multiple times through the backhaul links. Therefore, by serving the mobile users locally, edge-caching can jointly offload traffic burden on the backhaul links, reduce system costs and improve quality of service (QoS) of the mobile users.

Over the past few years, extensive research  has been carried out on edge-caching networks, which has mainly focused on the performance analysis of caching, cache placement optimization and transmission strategies. A cache placement algorithm has been proposed to minimize the excepted downloading time for contents in~\cite{shanmugam2013femtocaching}. In~\cite{peng2014joint}, physical layer features are used in the cache placement problem to minimize network cost while to satisfy users' QoS requirements. The authors in~\cite{vu2018edge} investigated energy efficiency and time delivery of an edge-caching network. In addition, various coding schemes, intra and inter sessions, have been proposed to enhance caching performance~\cite{shanmugam2013femtocaching,maddah2014fundamental,han2016phy}.

The main assumption of the aforementioned papers is that the content popularity is known in advance. However, in practice, the popularity is unknown and has to be estimated and predicted. In this respect, the popularity learning problem can be categorized in two general approaches: model-free and model-based. In the model-free approach, there is no assumption on the content request distribution. The popularity learning is then performed within the process of optimizing a reward function (e.g cache hit ratio) by the so-called exploration-exploitation procedure.  Multi-armed-bandit (MAB) and reinforcement learning algorithms are mostly based on this approach which also have been adapted to  edge-caching applications~\cite{muller2017context,song2017learning,sadeghi2018optimal,somuyiwa2018reinforcement}. On the other hand, in the model-based approach, it is assumed that the content requests are generated by a parametric distribution. The Poisson stochastic process is a popular model  adopted in the content delivery networks~\cite{garetto2016unified} and also has been used in  edge-caching~\cite{bharath2016learning}. Once the request is modeled, the next step is to estimate the popularity. A simple way  is to take the average of  instantaneous requests, which is equivalent to the maximum likelihood estimation (MLE) from the estimation theory perspective. However, the MLE suffers from overfitting especially in edge caching systems where only a few request observations are available. For example, as it is reported
in~\cite{paschos2016wireless}, a  base station cache typically may receive 0.1 requests/content/day which is too small in contrast with  a typical content delivery network cache which normally receives 50 requests/content/day.

To improve the popularity estimation accuracy, side information  (user profile and content features) can also be incorporated in learning algorithms. In ~\cite{bacstuug2015transfer,bharath2016learning}, user profiles
are leveraged to speed up the learning convergence rate. One
important issue with this kind of side information is that
users may not be willing to share their personal profiles to
the edge-cache entity. On the other hand, content features (e.g topic categories) can be easily and cheaply obtained from the content server without jeopardizing users' privacy. In addition, knowing the most important content features can be useful to design  advanced cache-placement algorithms. For example, the authors of~\cite{zeng2018temporal} observed that there  is  a traffic pattern under different topic categories of contents  by  doing experimental validation on the dataset of a real   mobile  network. Therefore,  besides learning the popularities of the contents, in order to have a better understanding about the hidden request pattern, it is advantageous to also learn the importance of content features.


 In this paper, we take the content features  into account and introduce a new probabilistic model for the content requests.   The learning process is performed in the Bayesian paradigm which is robust against overfitting and provides a way to quantify our uncertainty about the estimation. The model allows us to define  different types of predictive distributions by which we can effectively model the uncertainty of future requests. The statistical information of these posterior predictive distributions can be used to design a sophisticated caching policy. Here, we should also mention that the central contribution of this paper is not to devise  a caching policy but rather  to propose a more accurate and reasonable  probabilistic model for content requests.
Overall, the main contributions of the paper are summarized as:
 \begin{itemize}
 	\item We provide a probabilistic model, a Poisson regressor based on a Gaussian process, for stationary content requests which captures the similarity between contents. The Gaussian process is  a very flexible and powerful statistical model that can model  nonlinear relationships between the popularities and  the features.

 	\item  The parameters of the model are learnt  in the  Bayesian framework. Due to few request samples in the local cache, Bayesian learning provides a powerful framework to mitigate overfitting. 
 	\item For prediction,  two types of predictive distributions are specified. One is used to predict the future requests  for the existing contents and the other to predict the popularity of a new content that may come to the system.
 \end{itemize}

The rest of the paper is organized as follows: the system model  and problem statement are described in Section \ref{seq:Sysmodel}. In Section \ref{BayesinInfer}, we apply the Bayesian approach for popularity learning. Finally, Section \ref{SimRes} shows the simulation results and Section \ref{Conclusion} concludes the paper.

\section{System Model and problem formulation}\label{seq:Sysmodel}
In this paper, we consider a cellular network consisting of a base station (BS) serving its mobile users. Users can make random requests from
a library of contents ${\cal C} = \left\{ {{c_1},...,{c_M}} \right\}$, where $M$ is the total number of contents. Each content is  assumed to have a set of features. For instance, a video content may  have a specific topic (e.g education, entertainment, science-technology,.. )  and some other features such as release year. We use ${{\bf{x}}_m}$ to be the feature vector of content $c_m$ with $Q$ dimensions whose  values can be either binary or continuous.  

The BS is equipped with a  limited capacity cache memory, and is connected to the remote content server through the backhaul links. Additionally, the remote
server has access to the whole content library ${\cal C}$. At each time slot\footnote{The time slots can be hours, days, etc.}, each user independently requests a content (or contents)\footnote{There is no limitation on the  number of requests by a user at a time slot} from the library ${\cal C}$. To alleviate the traffic burden on the backhaul links and increase the users' QoS,  some contents  are stored in the cache depending on the caching policy. The requested contents by the users will be served directly if they are already cached; otherwise they are fetched from the content server.
We suppose that the cache  module of the BS can only monitor the number of user requests towards contents of the library and cannot perform any user profiling. In addition, it is assumed that the content popularity is  fixed (we can assume it does not change over short time intervals, e.g. a few days) and the requests  are samples generated from a stationary distribution.

We define ${{\bf{d}}_c}\left[ {{T_n}} \right] = \left[ {{d_{{c_1}}}\left[ {{T_n}} \right],...,{d_{{c_M}}}\left[ {{T_n}} \right]} \right]^T$ to be the request vector where $d_{{c_m}}\left[ {{T_n}} \right]$ is the total number of requests for content $m$ during time slot $n$ with duration $T_n$. For simplicity, we assume that $T_n = T_{n'}$, $\forall n' \ne n$ . Therefore, we can drop $T$ and show the request vector by ${{\bf{d}}_{c,n}} = \left[ {{d_{{c_1},n}},...,{d_{{c_M},n}}} \right]^T$.  Also,  the requests for $n' \ne n$   are  presumed to be statistically independent random variables.  A common parametric model for  the requests  is  the  Poison stochastic process and the MLE approach to estimate the  rate request, or the popularity (we use the terms rate and popularity interchangeably)~\cite{bharath2016learning} as:
\begin{equation}\label{MeanPoisson}
{r_m} = \frac{{\sum\limits_{n = 1}^N {{d_{{c_m},n}}} }}{N},\quad \forall m = 1,...,M
\end{equation}
where ${r_m}$ is the popularity of content $c_m$ and $N$ is the total number of  request observations during the training period. Although this approach is  simple, it is not very accurate for popularity estimation. Firstly, MLE suffers from severe overfitting especially when the training set has only a few request observations. Secondly, it cannot incorporate  any kind of side information. For example, users commonly request contents based on their features. Therefore, we expect content popularities to be correlated in the feature space. By appropriately using this underlying prior knowledge about requests, the accuracy of popularity estimation can be significantly improved. In the next sections, we present our probabilistic model in order to deal with these issues. Before introducing the model, we summarize  the basic concepts of Gaussian processes which are essential for the subsequent sections.
\subsection{Gaussian Process in a Nutshell}\label{seq:GP}
A Gaussian process is a collection of random variables, any finite number of which have a joint Gaussian distribution. Using a Gaussian process, we can define a distribution over functions $f\left( {\bf{x}} \right)$:
\begin{equation}\label{GPdef}
f\left( {\bf{x}} \right)\sim {\cal G}{\cal P}\left( {{{\mu }}\left( {\bf{x}} \right),{{K}}\left( {{\bf{x}},{\bf{x'}}} \right)} \right)
 \end{equation}
where $\bf x$ is an arbitrary  input\footnote{Here, input is a very general concept that can be time, location, ... .  In  our problem, it is content features.} variable with $Q$ dimensions, and the  mean function, ${{{\mu }}\left( {\bf{x}} \right)}$, and the Kernel function, $K\left( {{\bf{x}},{\bf{x'}}} \right)$,   are respectively defined as:
\begin{align}
&{\bf{\mu }}\left( {\bf{x}} \right) = E\left[ {f\left( {\bf{x}} \right)} \right]\\
&K\left( {{\bf{x}},{\bf{x'}}} \right) = E\left[ {\left( {f\left( {\bf{x}} \right) - \mu \left( {\bf{x}} \right)} \right)\left( {f\left( {{\bf{x'}}} \right) - \mu \left( {{\bf{x'}}} \right)} \right)} \right].
\end{align}
This means that any finite collection of function values has a joint Gaussian distribution:
\begin{equation}
\left[ {f\left( {{{\bf{x}}_1}} \right),...,f\left( {{{\bf{x}}_M}} \right)} \right]^T\sim {\cal N}\left( {{\boldsymbol{\mu }},{\bf{K}}} \right)
\end{equation}
where ${\boldsymbol{\mu }} = {\left[ {\mu \left( {{{\bf{x}}_1}} \right),...,\mu \left( {{{\bf{x}}_M}} \right)} \right]^T}$ and the covariance matrix ${\bf K}$ has the entities ${\left[ {\bf{K}} \right]_{i,j}} = K\left( {{{\bf{x}}_i},{{\bf{x}}_j}} \right)$.
The kernel function specifies the main characteristics of the function that we wish to model and the basic assumption is that  variables with inputs $\bf x$ which are close are likely to be correlated. Choosing a good kernel function  for a learning task depends on intuition and experience. 
  A popular and simple kernel    is the squared exponential kernel (SEK):
 \begin{equation}\label{eq:SEK}
 K\left( {{{\bf{x}}_i},{{\bf{x}}_j}} \right) = {\theta _1}{e^{ - \sum\limits_{q = 2}^{Q+1} {{\theta _q}{{\left\| {x_i^{(q-1)} - x_j^{(q-1)}} \right\|}^2}} }}
 \end{equation}
 where ${\theta _1}$ is the vertical scale variation and ${\theta _{q + 1}}$ is the horizontal scale variation on dimension $q$  of the function. By using  different scales for each input dimension, we let them to have different importance. If ${\theta _{q + 1}}$ is close to zero, dimension $q$ will have little influence on the  covariance of variables.  Covariance function \eqref{eq:SEK} is infinitely differentiable and is thus
 very smooth. More details about the Gaussian process and the kernel functions can be found in~\cite{rasmussen2004gaussian}.
 \subsection{The proposed model}\label{seq:propModel}
 In this subsection, we introduce our probabilistic model for content requests.
 The following  regression-based hierarchical (multilevel) probabilistic model is proposed:
 \begin{subequations}\label{eq:GPdef}
\begin{align}
&{d_{{c_m},n}}|{\lambda _m}\left( {{{\bf{x}}_m}} \right)\sim Poi\left( {{e^{{\lambda _m}\left( {{{\bf{x}}_m}} \right)}}} \right),\forall n = 1,...,N \label{eq:Genlev1}\\
&{\lambda _m}\left( {{{\bf{x}}_m}} \right)|f\left( {{{\bf{x}}_m}} \right),{\theta _0}\sim {\cal N}\left( {f\left( {{{\bf{x}}_m}} \right),{\theta _0}} \right)\label{eq:Genlev2}\\
&f\left( {\bf{x}} \right)|{\bf{x}},{\theta _1},...,{\theta _{Q+1}}\sim {\cal GP}\left( {0,K\left( {{\bf{x}},{\bf{x'}}} \right)} \right). \label{eq:Genlev3}
\end{align}
\end{subequations}
The first level of the model, \eqref{eq:Genlev1},  is the Poisson observation   distribution for content requests.  At this level, the request for content $c_m$   is assumed to follow a Poisson distribution with  natural parameter ${\lambda _m}\left( {{{\bf{x}}_m}} \right)$ which  is a function of its features. We note that the request rate  is an exponential function of the natural parameter, ${r_m}\left( {{{\bf{x}}_m}} \right) = {e^{{\lambda _m}\left( {{{\bf{x}}_m}} \right)}}$. As we previously mentioned, it is expected that there is a similar request pattern  between contents with similar features. This prior information is employed at the higher levels.
 In \eqref{eq:Genlev2}, ${\lambda _m}\left( {{{\bf{x}}_m}} \right)$ follows a normal distribution with mean $f\left( {{{\bf{x}}_m}} \right)$ and variance ${\theta _0}$. By this assumption, we allow contents with exactly the same features to have different popularities which is possible in practice. At the higher level of the model, \eqref{eq:Genlev3}, we assume that $\left\{ {f\left( {{{\bf{x}}_m}} \right)} \right\}_{m = 1}^M$ are realizations of function $f\left( {\bf{x}} \right)$ drawn from a Gaussian process with zero mean and kernel function $ K$. By this assumption, contents with similar features are encouraged to be correlated in the feature space. 
\section{Bayesian Learning}\label{BayesinInfer}
\subsection{Inference}
In this section, we exploit the Bayesian framework to learn the probabilistic model in \eqref{eq:GPdef}. In other words, given the content request observations ${\cal D} = \left\{ {{{\bf{d}}_{c,n}},...,{{\bf{d}}_{c,n}}} \right\}_{n = 1}^N$, we aim to update our belief  about the  model's parameters $\left\{ {{\lambda _m}\left( {{{\bf{x}}_m}} \right)} \right\}_{m = 1}^M,f\left( {\bf{x}} \right)$. However, we cannot estimate the infinite-dimensional function $f\left( {\bf{x}} \right)$ and hence the focus is only on the realizations at  $\left\{ {f\left( {{{\bf{x}}_m}} \right)} \right\}_{m = 1}^M$.
Moreover, to simplify the inference, we  can integrate out  ${f\left( {{{\bf{x}}_m}} \right)}$ from the model. By doing this, we have: 
\begin{equation}
{\boldsymbol{\lambda }} = {\left[ {{\lambda _1}\left( {{{\bf{x}}_1}} \right),....,{\lambda _M}\left( {{{\bf{x}}_M}} \right)} \right]^T}\sim {\cal N}\left( {{\bf{0}},{\bf{\tilde K}}} \right)
\end{equation}
where ${\bf{\tilde K}} = {\bf{K}} + {\theta _0}{\bf{I}}$.
Additionally, in practice, the available prior knowledge may not be enough to fix the  parameters $\left\{ {{\theta _q}} \right\}_{q = 0}^{Q+1}$ . A common approach to estimate these parameters is cross validation. However, this trial and error experiment may be  tedious and computationally extensive. A very systematic way to learn these parameters is to model their uncertainty by a prior distribution. Since the values of ${\theta _0},..,{\theta _Q}$ must be positive, a natural choice would be  Gamma priors:
\begin{align}\label{KernlHyper}
{\theta _q}\sim Gam\left( {{A_q},{B_q}} \right) \; \; \forall q =0,...,Q+1
\end{align}
where ${A_q}$ and ${B_q}$ are respectively the shape and the rate of each Gamma distribution.
\begin{figure}
	\centering
	\includegraphics[width=2in]{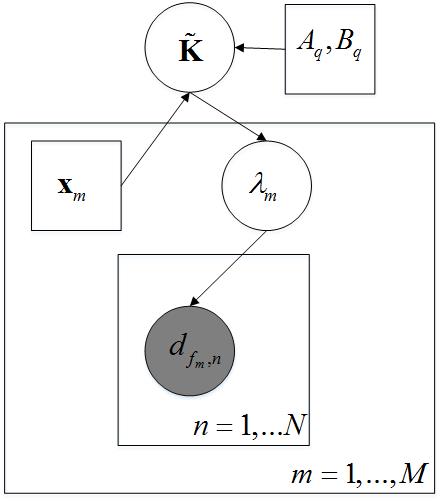}
	\caption{The proposed probabilistic model for content requests}
	\label{fig_simllg}
\end{figure}

Fig.\ref{fig_simllg} shows the graphical representation of the Bayesian model. The shaded node represents the observed requests and the plates represent multiple samples of random variables. The unshaded circle nodes indicate unknown quantities and the squares show the deterministic parameters of the model.

The inference of all unknown variables of the  model is given by the Bayes rule as:
\begin{align}\label{eq:BayesRule}
p\left( {{\boldsymbol{\lambda }},\left\{ {{\theta _q}} \right\}_{q = 0}^{Q+1}|{\cal D}} \right) = \frac{{\prod\limits_{n = 1}^N {\prod\limits_{m = 1}^M {p\left( {{d_{{c_m},n}}|{\lambda _m}} \right)} } p\left( {{\boldsymbol{\lambda }}|{\bf{\tilde K}}} \right)\prod\limits_{q = 0}^{Q+1} {p\left( {{\theta _q}} \right)} }}{Z}
\end{align}
where $p\left( {{\boldsymbol{\lambda }},\left\{ {{\theta _q}} \right\}_{q = 0}^{Q+1}|{\cal D}} \right)$ is the posterior distribution and
 the   denominator $Z$ is a normalization constant. 
 Unfortunately, the normalization constant is intractable to compute and there is no closed-form expression for the posterior distribution.
So, instead, we use a Monte Carlo Markov Chain (MCMC) method to approximate the posterior distribution. Specifically, we use the Hamiltonian Monte Carlo (HMC) method which has been one of the most successful MCMC methods to sample from an unnormalized distribution. Now, we  give an overview of the HMC. The complete description can be found in~\cite{neal2011mcmc}.

HMC is based on the simulation of Hamiltonian dynamics as a method to generate a sequence of samples $\left\{ {{{\boldsymbol{\zeta }}_s}} \right\}_{s = 1}^S$  from a desired  $D$-variate distribution $p\left( {\boldsymbol{\zeta }} \right)$ by exploring its sample space.
 It combines gradient information of $p\left( {\boldsymbol{\zeta }} \right)$ and auxiliary variables, ${\bf p} \in R^{D \times 1} $, with density $p\left( {\bf{p}} \right) = {\cal N}\left( {{\bf{0}},{\bf{G}}} \right)$.   The Hamiltonian function is then defined as: 
\begin{equation}\label{JointLLgHam}
H\left( {{\boldsymbol{\zeta }},{\bf{p}}} \right) =   \psi \left( {\boldsymbol{\zeta }} \right) + \frac{1}{2}\log {\left( {2\pi } \right)^D}{\bf{G}} + \frac{1}{2}{{\bf{p}}^T}{\bf{G}}{\bf{p}}
\end{equation}
where $\psi \left( \boldsymbol \zeta  \right)$ is the negative log of the unnormalized $p\left( \boldsymbol \zeta \right)$ and  ${\bf{G}}$
is usually assumed to be the identity matrix. The physical analogy of \eqref{JointLLgHam} is the Hamiltonian dynamics which describe the sum of the potential energy (the first term) and the kinetic energy (the last two terms).

Hamiltonian dynamics are simulated by discretizing their continuous analogue equations using the leapfrog method. This discretization has two parameters, number of leapfrog steps $L$ and step-size $\varepsilon$. The full description of a movement in HMC which is from a current state (sample) to a new state is depicted in Alg.\ref{algHMC1}. 
HMC is only applicable for differentiable and unconstrained variables. However, in \eqref{eq:BayesRule}, there are some variables, ${\left\{ {{\theta _q}} \right\}_{q = 0}^{Q+1}}$,  that must be positive. To handle this issue, we exploit the exponential-transformation where instead of $\theta _q$, we use ${\phi _q}=\log (\theta _q)$ with ${{\phi _q}}$ serving as an unconstrained auxiliary variable. Note that to use these transformations, we also need  to compute the Jacobian determinant as a result of the change of random variables.

By defining ${\boldsymbol{\zeta }} \!\!=\!\! {\left[ {{{\boldsymbol{\lambda }}^T},{\phi _0},...{\phi _{Q+1}}} \right]^T} \!\!\in {R^{(M + Q + 2) \times 1}}$ and  $p \left( \boldsymbol \zeta  \right)$ as the posterior distribution \eqref{eq:BayesRule}, the  negative log of  unnormalized $p \left( \boldsymbol \zeta  \right)$ (after the exponential-transformation)  is given by:
\begin{IEEEeqnarray}{rCl}\label{LogPost}
&\psi \left( \boldsymbol \zeta  \right)=-\log p\left( {{\boldsymbol{\lambda }},\left\{ {{\theta _q}} \right\}_{q = 0}^{Q+1}|{\cal D}} \right)\!\! =\!\! \sum\limits_{m = 1}^M {\sum\limits_{n = 1}^N { - {d_{c_mn}}{\lambda _m}} } 
+ {e^{{\lambda _m}}} \nonumber\\
&+ \frac{1}{2}\log \det \left( {{\bf{\tilde K}}} \right)
+\frac{1}{2} {{\boldsymbol{\lambda }}^T}{{{\bf{\tilde K}}}^{ - 1}}{\boldsymbol{\lambda }} + \sum\limits_{q = 0}^{Q+1} { - {A_q}{\phi _q} + {B_q}{e^{{\phi _q}}}}. \IEEEeqnarraynumspace
\end{IEEEeqnarray}
Also, the gradient of \eqref{LogPost}, which is required in Alg.\ref{algHMC1}, can be easily computed by using matrix derivatives~\cite{petersen2008matrix}:
\[\begin{array}{*{20}{l}}
{\frac{{\psi \left( {\boldsymbol{\zeta }} \right)}}{{\partial {\lambda _m}}} = \sum\limits_{n = 1}^N { - {d_{c_mn}}}  + N{e^{{\lambda _m}}} + {{\left[ {{{{\bf{\tilde K}}}^{ - 1}}{\boldsymbol{\lambda }}} \right]}_m}}\\
{\frac{{\psi \left( {\boldsymbol{\zeta }} \right)}}{{\partial {\phi _q}}} = \frac{1}{2}tr\left( {{{{\bf{\tilde K}}}^{ - 1}}\frac{{\partial {\bf{\tilde K}}}}{{\partial {\phi _q}}}} \right) - \frac{1}{2}{{\boldsymbol{\lambda }}^T}{{{\bf{\tilde K}}}^{ - 1}}\frac{{\partial {\bf{\tilde K}}}}{{\partial {\phi _q}}}{{{\bf{\tilde K}}}^{ - 1}}{\boldsymbol{\lambda }} - {A_q} + {B_q}{e^{{\phi _q}}}}.
\end{array}\]

\begin{algorithm}
	\SetAlgoLined
	\KwIn{$\boldsymbol \zeta _s ,\varepsilon ,L,{\nabla _{\boldsymbol{\zeta }}}\psi \left( {{\boldsymbol{\zeta }},} \right),\psi \left( {{\boldsymbol{\zeta }},} \right),{\bf{G}}$}
	\KwOut{ ${\boldsymbol \zeta}_{s+1}$ }
	\tcc{draw a  sample from $p\left( \zeta  \right)$ }
	
	${\bf q}_1={\boldsymbol \zeta}_{s}$, ${\bf{p}}_1\sim {\cal N}\left( {{\bf{0}},{\bf{G}}} \right)$\;
	Compute  $H\left( {{\boldsymbol q}_1 ,{\boldsymbol p}_1} \right)$\;
\For{$l \leftarrow 1$ \KwTo $L$}{
	${\bf{p}} \leftarrow {\bf{p}}_l - \varepsilon \nabla \psi \left( {\boldsymbol{q }}_l \right)$\;

	${\boldsymbol{q }}_{l+1} = {\boldsymbol{q }}_l + \varepsilon {{\bf{G}}^{ - 1}}{\bf{p}}$\;
	${\bf{p}}_{l+1} = {\bf{p}} - \varepsilon \nabla \psi \left( {\boldsymbol{q }}_{l+1} \right)$\;
}
compute $dH = H\left( {\boldsymbol{q }_{L+1},{\bf p}_{L+1}} \right) - H\left( { \boldsymbol q_1 ,{\bf {p}}_1} \right)$\;
 \eIf{$rand\left( {} \right) < {e^{-dH}}$}{
$\boldsymbol \zeta_{s+1}  = {\boldsymbol q }_{L+1}$\tcc*{accept }
}{
	$\boldsymbol \zeta_{s+1}   = {\boldsymbol q }_1$\tcc*{reject }
}
	\caption{The HMC sampling algorithm ~\cite{neal2011mcmc}}
	\label{algHMC1}
\end{algorithm}
 Once, we collect enough samples from the HMC, any function of the posterior distribution moments can be computed.
 The initial MCMC samples are  usually discarded because they may be far away from the true distribution. These samples are  called burn-in samples. 
 
 Nevertheless, our goal is not just to learn the parameters of the model based on the training set but is to make prediction about the possible content request values in future. The next subsection explains how this can be performed.
\subsection{Prediction}
Here, we aim to perform prediction in two ways. The first one is to predict the requests for the existing contents. This can be performed using the posterior predictive distribution (distribution of a new request) given in \eqref{pospred1}:
\begin{equation}\label{pospred1}
p\left( {{\bf{d}}_c^{new}|{\cal D}} \right) = \int {p\left( {{\bf{d}}_c^{new}|{\boldsymbol{\lambda }}} \right)p\left( {{\boldsymbol{\lambda }}|{\cal D}} \right)d{\boldsymbol{\lambda }}} 
\end{equation}
where ${p( {{\bf{d}}_c^{new}|{\boldsymbol{\lambda }}} )}$ is a Poisson distribution  and ${p\left( {{\boldsymbol{\lambda }}|{\cal D}} \right)}$ is the marginal posterior distribution of ${\boldsymbol{\lambda }}$. 
However, we would like to make a point prediction rather than dealing with the whole predictive distribution.  The best guess for a point estimation in the Bayesian context is based on risk (or loss) minimization~\cite[Chapter~2]{robert2007bayesian}. In other words, a loss function is defined which specifies the loss incurred by guessing the value ${{\bf{d}}^{new}}$ when the actual value is ${{\bf{d}}^*}$. The most common loss evaluation metric is the quadratic loss. The value of ${{\bf{d}}^{new}}$ that minimizes this risk function is the mean of the predictive distribution which can be approximated as:
\begin{equation}\label{mean_app}
E\left\{ {{\bf{d}}_c^{new}|{\cal D}} \right\} \approx  \frac{1}{S}\sum\limits_{s = 1}^S {{e^{{{\boldsymbol{\lambda }}_s}}}} .
\end{equation}

The second prediction task is to predict the popularity of a newly-added content that may enter the system. This can be calculated by a second type of posterior predictive distribution defined as:
\begin{equation}\label{eq:pospredtype2}
p\left( {{\lambda _{M + 1}}|{{\bf{x}}_{M + 1}}} \right) = \int {p\left( {{\lambda _{M + 1}}|{\boldsymbol{\lambda }},{\boldsymbol{\theta }},{{\bf{x}}_{M + 1}}} \right)p\left( {{\boldsymbol{\lambda }},{\boldsymbol{\theta }}|{\cal D}} \right)d{\boldsymbol{\lambda }}d{\boldsymbol{\theta }}}  
\end{equation}
where ${{\bf{x}}_{M + 1}}$ is the feature vector of the new content. To compute
${p\left( {{\lambda _{M + 1}}|{\boldsymbol{\lambda }},{{\bf{x}}_{M + 1}}} \right)}$, we note that the   joint distribution of $p\left( {{\lambda _1},...,{\lambda _{M + 1}}} \right)$ is a Normal distribution with zero mean and covariance matrix:
\[\left( {\begin{array}{*{20}{c}}
	{{\bf{\tilde K}}}&{{\bf{\tilde k}}}\\
	{{{{\bf{\tilde k}}}^T}}&{K\left( {{{\bf{x}}_{M + 1}},{{\bf{x}}_{M + 1}}} \right) + {\theta _0}}
	\end{array}} \right)\]
 where ${\bf{\tilde k}} = \left[ {K\left( {{{\bf{x}}_1},{{\bf{x}}_{M + 1}}} \right),...,K\left( {{{\bf{x}}_M},{{\bf{x}}_{M + 1}}} \right)} \right]{^T}$.
Based on the properties of Normal distributions, the conditional distribution ${p\left( {{\lambda _{M + 1}}|{\boldsymbol{\lambda }},{{\bf{x}}_{M + 1}}} \right)}$ is a Normal distribution with mean and variance:
 \begin{align}
 &{{\hat \lambda }_{M + 1}}\! =\! {{{\bf{\tilde k}}}^T}{{{\bf{\tilde K}}}^{ - 1}}{\boldsymbol{\lambda }}\nonumber \\
 &{{\hat \sigma }_{M + 1}} = K\left( {{{\bf{x}}_{M + 1}},{{\bf{x}}_{M + 1}}} \right) + {\theta _0} - {{{\bf{\tilde k}}}^T}{{{\bf{\tilde K}}}^{ - 1}}{\bf{\tilde k}}\nonumber.
 \end{align}
 Again, the  optimal predictive value for \eqref{eq:pospredtype2} considering the quadratic loss is its mean. It should be noted that \eqref{eq:pospredtype2} is the distribution of the natural parameter of a new content. The point estimation of the request rate   can be approximated as:
 \begin{equation}\label{eq:mean_apptyep2}
E\left( {{r _{M + 1}}|{{\bf{x}}_{M + 1}}} \right)  \approx  \frac{1}{S}\sum\limits_{s = 1}^S {{e^{{\bf{\tilde k}}_s^T\left( {{{\bf{x}}_{M + 1}},{\bf{x}}} \right){\bf{\tilde K}}_s^{ - 1}{{\boldsymbol{\lambda }}_s}}}}. \end{equation}

\section{Simulation Results}\label{SimRes}
In this section, we present our simulation results to show the performance of the proposed probabilistic content request model denoted by "\textit{Bayesian Poisson-GP}". To compare our results, we use the independent Poisson model with MLE in \eqref{MeanPoisson} denoted by "\textit{MLE Poisson}"  as a benchmark. As far as the HMC technique is concerned, we set $\varepsilon =.015$ and $L=20$ and ran it for 5000 samples where the first 2500 samples were considered as the burn-in samples. The number of features is $Q=4$ and specifically features ${x}_m^{(1)}$, ${x}_m^{(2)}$, ${x}_m^{(3)}$ are binary whose values are randomly generated from Bernoulli distributions with parameters $0.5$, $0.8$ and $0.2$ for all $m$, respectively. Feature $x_m^{(4)}$ is continuous and generated from a Normal distribution with zero mean and unit variance for all $m$. Moreover, the parameters of the Kernel function \eqref{eq:SEK} are ${\theta _0} = .0001,{\theta _1} = 0.1,{\theta _2} = 0.25,{\theta _3} = 0,{\theta _4} = 0.1$ and ${\theta _5} = 0.5$.

\begin{figure}
	\centering
	\includegraphics[width=7.5cm, height=4.5cm,trim={33 0 0 35}]{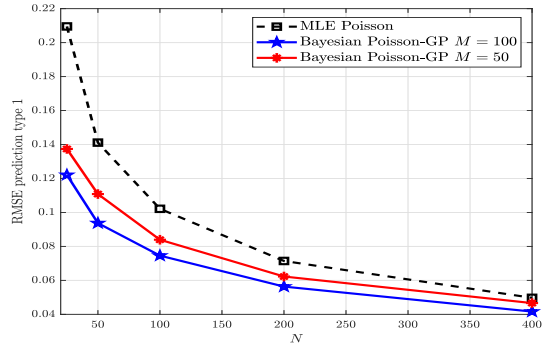}
	\caption{RMSE prediction type 1 versus $N$ request observations}
	\label{fig:RMSEtype1}
\end{figure}

Fig.\ref{fig:RMSEtype1} shows the root mean square error (RMSE) of the popularity predictive type 1 in \eqref{mean_app} versus the number of observations in the training set, $N$. It can be seen that the Bayesian Poisson-GP  significantly performs better than the MLE Poisson. We can also observe that as the number of contents increases, the Bayesian Poisson-GP performance is improved. This is because as $M$ increases, the Gaussian process can  learn better the relationship between the popularities and the features.

\begin{figure}
	\centering
	\includegraphics[width=7.5cm, height=4.5cm,trim={33 0 0 30}]{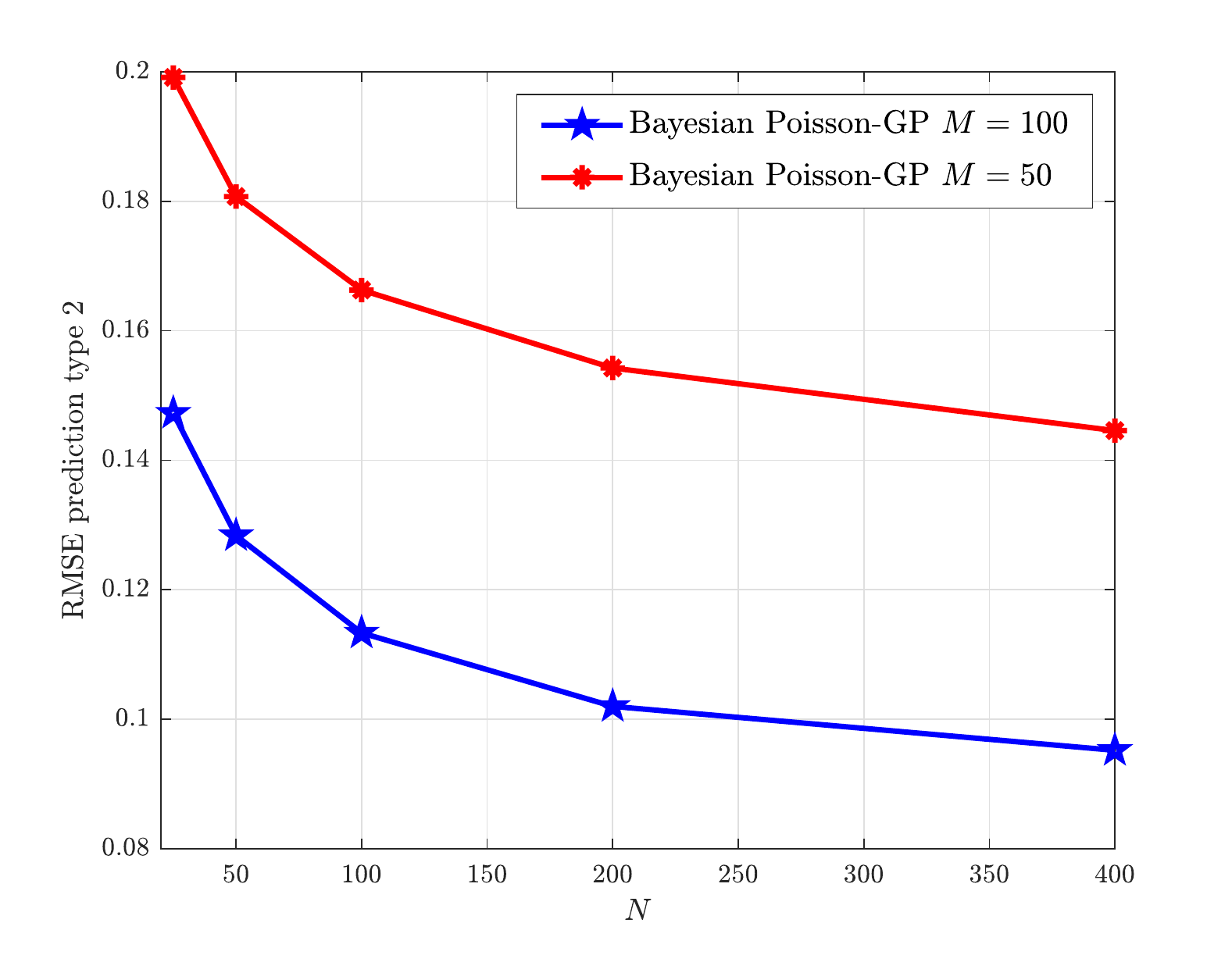}
	\caption{RMSE prediction type 2 versus $N$ request observations}
	\label{fig:RMSEtype2}
\end{figure}

Now, we investigate the performance of our model in terms of how well it can predict the popularity of a new content (popularity predictive type 2 in \eqref{eq:mean_apptyep2}). The feature of the new content is randomly generated with the same process as for the existing contents. Fig. \ref{fig:RMSEtype2} shows the RMSE  of the predicted popularity  of the new content versus $N$. As we see the performance of the model improves when the size of the training set or the number of contents increase. In this scenario, there is no explicit way  to use the content features in the MLE Poisson to make prediction about the popularity of a new content, therefore we are unable to compare  our model with it.

Next, we show the accuracy of Kernel parameter learning efficiency of our model. Tables \ref{table1} and \ref{table2} show the estimated mean values of the kernel function parameters. As we expected, it is observed that as the number of observations increases we get closer to the true values. However,  from the tables, the accuracy improvement of the parameters is largely affected by the number of contents.  For example, for feature $x_m^{(2)}$, which does not affect the outcome of the model, the value of its scale variation, ${\theta _3}$, has a better estimation at $N=400$ for $M=100$ in comparison with $M=50$. These results confirm our previous simulations that  as $M$ increases the Gaussian process gets more accurate and consequently shows a better prediction performance. The reason for this behavior is that by increasing $M$ the number of observations in the feature space increases which results in a better prediction accuracy.

\begin{table}
	\centering
	\begin{tabular}{ |c|c|c|c|c|c|c| } 
		\hline
		&True value& N=25 & N=50& N=100& N=200& N=400 \\ 
		\hline
		${\theta _0}$&0.0001&0.0081&0.0032 &   0.0031  &  0.0016 &0.0009    \\
		\hline
		${\theta _1}$ &0.1&0.1187&    0.1443&    0.1268 &   0.1313 &   0.1269   \\
		\hline
		${\theta _2}$ &0.25& 0.1633&    0.1553 &   0.1879 &   0.1848&    0.2083 \\
		\hline
		
		${\theta _3}$ &0& 0.0676&    0.0484 &   0.0383 &   0.0146 &   0.0192  \\
		\hline
		${\theta _4}$&0.1&  0.0755    &0.0535 &   0.0542  &  0.0871  &  0.0843  \\
		\hline
		${\theta _5}$ &0.5&0.3354&    0.3441&    0.3904 &   0.4180&    0.4495   \\
		\hline
		
	\end{tabular}
	\caption{the value of estimated kernel function parameters for $M =50$}
	\label{table1}
\end{table}

\begin{table}
	\centering
	\begin{tabular}{ |c|c|c|c|c|c|c| } 
		\hline
		 & True value&N=25 & N=50& N=100& N=200& N=400 \\ 
		\hline
		${\theta _0}$ &0.0001&0.0035&    0.0014&    0.0010 &   0.0006 &   0.0002  \\
	\hline
		${\theta _1}$ &0.1&0.1179&    0.1225&    0.1141 &   0.1117&    0.1129    \\
	\hline
	${\theta _2}$ &0.25&0.2187&    0.2296 &   0.2232 &   0.2451  &  0.2428  \\
	\hline

	${\theta _3}$ &0&0.0466&    0.0179 &   0.0072 &   0.0077  &  0.0045   \\
	\hline
	${\theta _4}$ &0.1&0.0736&    0.0762 &   0.0902 &   0.0969  &  0.1043  \\
		\hline
		${\theta _5}$ &0.5&0.3732&    0.4504 &   0.4649 &   0.4536&    0.4753 \\
			\hline
		
	\end{tabular}
		\caption{the value of estimated kernel function parameters for $M =100$}
		\label{table2}
\end{table}

\section{Conclusions}\label{Conclusion}
In this paper, we proposed a flexible model for modeling the content requests and predicting their popularity. We proposed a multilevel probabilistic model, the Poisson regressor based on Gaussian process,   that can capture the similarity between contents in terms of their features. We utilized  Bayesian learning to obtain the parameters of the  model because  it is robust against overfitting and therefore  efficient in edge-caching system where overfitting is a big challenge due to small number of request observations. Then, two posterior predictive distributions were specified for prediction purposes.   In the simulation results, we showed that the Bayesian Poisson-Gaussian process structure significantly outperforms the MLE independent Poisson in terms of content popularity prediction. 
 


\section*{Acknowledgment}

This work was funded by the National Research Fund (FNR),
Luxembourg under the projects "LISTEN" and "PROCAST". This work was also supported  by the European Research Council (ERC) under the project "AGNOSTIC".


%

\bibliographystyle{IEEEtran}

\end{document}